# Generalized Thermostatistics and Wavelet Analysis

# of the Solar Wind and Proton Density Variability

Maurício José Alves Bolzan[1,*], Reinaldo Roberto Rosa[2], Fernando Manuel Ramos[2], Paulo Roberto Fagundes[1], Yogeshwar Sahai[1]

1. Instituto de Pesquisa e Desenvolvimento, Universidade do Vale do Paraíba, São José dos Campos, Brazil

2. Laboratório Associado de Computação e Matemática Aplicada, Instituto Nacional de Pesquisas Espaciais, São José dos Campos, Brazil

*. Corresponding Author: Maurício J. A. Bolzan (bolzan@univap.br), Fax: (12)39471149, Av. Shishima Hifumi, 2.911 – Urbanova, CEP 12244-000 - São José dos Campos, SP, Brazil.

**Abstract**

In this paper, we analyze the probability density function (PDF) of solar wind velocity and proton density, based on generalized thermostatistics (GT) approach, comparing theoretical results with observational data. The time series analyzed were obtained from the SOHO satellite mission where measurements were sampled every hour. We present in the investigations data for two years of different solar activity: (a) moderate activity (MA) period (1997) and (b) high activity (HA) period (2000). For the MA period, the results show good agreement between experimental data and GT model. For the HA period, the agreement between experimental and theoretical PDFs was fairly good, but some distortions were observed, probably due to intermittent characteristics of turbulent processes. As a complementary analysis, the Global Wavelet Spectrum (GWS) was obtained allowing the characterization of the predominant temporal variability scales for both the periods and the stochastics aspects of the nonlinear solar wind variability are discussed.

# Introduction

Due to the important role of the solar wind properties in the solar-terrestrial plasma relations and magnetospheric physics, the study of its statistical properties and their relations to those in the geomagnetic indices has been attracting growing interest (Kovács et al., 2001; Lui, 2002; Hnat et al., 2002). Statistical behaviour of velocity field fluctuations recorded in wind tunnels and these obtained from solar wind observations exhibit striking similarities (Hnat et al., 2002), where a common feature found in both fluctuations is the presence of statistical intermittency (Burlaga, 1991; Marsch and Tu, 1994; Marsch and Tu, 1997; Burlaga and Foreman, 2002). The intermittency phenomena in the framework of turbulence theory has been investigated by many authors through laboratory and numerical experiments (e.g. Anselmet et al., 1984; Frisch, 1995; Ramos et al., 2001a) and the investigation of turbulent hydrodynamical flows has been developed considering many different approaches: Reynolds-stress models; subgrid-scale models for large-eddy simulations (LES); spectral models; and probability density functions (PDF) models (Frisch, 1995). On the other hand, the turbulence modeling of intermittent magnetohydrodynamical (MHD) flows are based on: (i) the She-Leveque approach, that describes the observed scaling structure function (Biskamp and Müller, 2000); (ii) Fokker-Planck equation considering the Castaing distribution (Castaing et al., 1990; Hnat et al., 2002). Such approaches are used in the isotropic inertial subrange of turbulent fluctuations assuming the Kolmogorov hypothesis and contain the energy cascade phenomenology (Kolmogorov, 1941, 1962).

Usually, the properties of turbulent flows, despite the nature of the underlying physics, are studied from the probability density functions (PDFs) of fluctuating quantities (velocity differences, for example) at different separation scales. It is a well-known property of turbulent flows that, at large scales, these PDFs are normally distributed. However, at increasingly smaller scales, they become strongly non-Gaussian and display tails flatter than expected for a Gaussian process. This is interpreted as the signature of the intermittency: the emergence of strong bursts in the kinetic energy dissipation rate. Within this framework, intermittency and non-extensive turbulence are linked by the entropic parameter q from generalized thermostatistics theory (Ramos et al., 1999; Beck, 2000; Arimitsu and Arimitsu, 2000; Ramos et al, 2001a, 2001b, Bolzan et al., 2002, Ramos et al. 2004). Recently, characterization of intermittent turbulence in the solar wind velocity was performed using generalized distribution from the nonextensive statistics approach (Burlaga and Viñas, 2004). The nonextensive parameter q represents a measurable quantity, flow independent and robust to variations in the Reynolds number, that can be used to quantify the ocurrence of intermittency in turbulent flows. Moreover, the existence of possible coherent structures in the time-frequency domain related to intermittent turbulent fluctuations can be well characterized by means of the global wavelet spectra (Rosa et al., 2002). In this paper, analysing the solar wind and proton density data, we show that this new approach provides interesting insights on MHD turbulence in a complex environment such as the solar-terrestrial plasma.

# Data

The data of solar wind and proton density analysed in this work were observed by the SOHO satellite and provided by the University of Maryland data base (http://umtof.umd.edu/pm/crn/). The time series used were measured in the years 1997 and 2000 and were sampled at the rate of 1 measurement per hour. Figure 1 shows both the time series (solar wind velocity and proton density) for 12 months of 1997 and Figure 2 shows the similar data sets for 12 months of 2000.

The solar wind for 1997 has moderate amplitude when compared with the similar data set for 2000. The plot of solar wind for 2000 represents a characteristic scenario in the evolution of the solar cycle, notably the impulsive change of the velocity amplitude in the month July. This enhancement in the solar wind velocity is associated with the very strong solar disturbance that occurred in July. Also, for the proton density time-series, for both the MA and HA periods, we found significant differences in temporal variability, notably a higher amplitude for 2000 compared with 1997. It is important to point out that the usage of proton density data has larger errors than the velocity data, which is an intrinsic problem in the analysis of the data, however there is some consistency in the proton density and velocity data used here.

## Theory

Recently, Kovács et al. (2001) have suggested that, disregarding storms and/or substorms as the main sources of the evolution of the geomagnetic disturbances, the fluctuating nature of the field can be interpreted in the present context as manifestation of turbulent phenomena that take place within the plasma of the magnetosphere. It has long been accepted that turbulence evolves through cascade processes that involve a hierarchy of coherent vortex structures belonging to a wide range of spatial scales. Kolmogorov (1962) proposed the inhomogeneous flow down (cascade) of the energy from system-size scales to dissipative ones. The inhomogeneity involves the singular behaviour of the energy distribution in physical space resulting in strong gradients, or intermittency in the time-series of the energy related physical quantities of the system, e.g. velocity (Ramos et al., 2001a), temperature (Bolzan et al., 2002) or magnetic fields (Kovács et al., 2001; Lui, 2002).

In this paper, we will adopt a generalization of the PDF model used in our previous works (Ramos et al., 1999; Ramos et al., 2001a and b; Bolzan et al., 2002), assuming that $p_q(v_r)$ is given by:

$$p_q(v_r) = \left[1 - \beta(1-q)\left(|v_r|^{2\alpha} - C\, sign(v_r)\left(|v_r|^{\alpha} - \frac{1}{3}|v_r|^{3\alpha}\right)\right)\right]^{1/(1-q)} / Z_q \qquad (1)$$

where *C* is a small skewness correction term, and $Z_q$ is given by

$$Z_q = \left\langle |v_r|^0 \right\rangle = \frac{a^{m_0+1}}{\alpha} B(\phi_0, \chi_0) \qquad (2)$$

with $B(\phi_0,\chi_0) = \Gamma(\phi_0)\, \Gamma(\chi_0)\, /\, \Gamma(\phi_0+\chi_0)$, $\phi_0 = (1 + m_0)/2$, $\chi_0 = 1 - \phi_0$, $l = 1/(q-1)$, $m_0 = (1-\alpha)/\alpha$, and $a = \sqrt{l/\beta}$.

Neglecting the skewness correction term, we obtain for the PDF n-th moment:

$$\left\langle |v_r|^n \right\rangle = a^{m_n - m_0} \frac{B(\phi_n,\chi_n)}{B(\phi_0,\chi_0)} \tag{3}$$

where $\phi_n = (1 + m_n)/2$, $\chi_n = l - \phi_n$ and $m_n = \dfrac{(n+1) - \alpha}{\alpha}$.

The parameters q and β determine the shape of the PDF and are computed from equation 3 (Bolzan et al., 2002; Ramos et al., 2004). Thus, note that the q and β parameters are derived from the experimental kurtosis for each scale.

We also used the Global Wavelet Spectrum (GWS) through the Morlet Wavelet Transform (MWT). This mathematical tool is similar to Power Spectrum Density (PSD) obtained by Fast Fourier Transform (FFT), and is based in the calculation of variance in each scale, or period, obtained by MWT. The objective of this procedure is to identify the predominant scales (periods) driving the turbulent process. For this, the computation consists in to sum all energy associated with each scale. This can be performed according to the following equation (Le and Wang, 2003):

$$M(a) = \int |W(a,t)|^2 \, dt , \qquad (4)$$

where *a* is the scale, W(*a*,t) is the Morlet wavelet transform applied in the time-series, and t is the temporal size of the time-series.

## Results and Discussions

In order to validate the model described in section 2, we compared measured distributions, corresponding to two different periods of solar activity, for both variables, solar wind and proton density, with the theoretical PDFs obtained from equation 1. For each data set, we measured the variance and kurtosis, which allowed us to compute q and β by means of the corresponding expressions obtained from equation 3. The parameter α was chosen according to the empirical formula α = 6 - 5q as used by Bolzan et al. (2002) and Ramos et al. (2004).

Figure 3 presents the theoretical and experimental semi logarithmic plots of $p_q(x_r)$ versus $x_r$ at four different scales, properly rescaled and vertically shifted for better visualization, for the solar wind velocity in 1997. The increment scales r used were $r = (2, 20, 200, 2000)$ and correspond to lags of $r_1 = 2$ hours, $r_2 = 20$ hours, $r_3 = 8.3$ days and $r_4 = 83.3$ days. These scales are similar to the ones used by Burlaga and Viñas (2004). Overall, we observe that the theoretical results (solid lines) are in good agreement with measurements across spatial scales spanning three orders of magnitude and a range of up to 5 standard deviations, including the rare fluctuations at the tail of the distributions. We performed a simple error analysis given by the correlation coefficient between experimental and theoretical PDF for each scale, as shown in Table 1. We note that there are higher values of correlation coefficient for all the increment scales, indicaty good agreement between experimental data and our model results. The transition from large-scale Gaussian behavior to a stretched exponential form as r decreases is quite evident and well reproduced by Tsallis´ distribution (Tsallis, 1988). At small scales, the distributions have tails larger than that expected for a normal process. This excess of large fluctuations, compared to a Gaussian distribution, is a well known signature of intermittency. The spiky shape near the origin is also a signature of intermittency (Frisch, 1995). According to Burlaga and Viñas (2004), these high points on the tail of the distribution in PDF for $r_1 = 2$ hours, represents a few large jumps in the solar wind velocity associated with shocks, stream interfaces, and some discontinuities with large shear. Furthermore, PDF for $r_2 = 20$ hours presented significative positive skewness, similar to results reported by Burlaga and Viñas (2004) for lags of 16 hours. According to these authours, this skewness is a consequence of stream steepening, i.e., faster plasma overtaking slower plasma. The PDF for $r_3 = 8.3$ days, presents a transition between the stretched exponential to Gaussian form. This behavior may be associated with the slow flows, that have temporal scales in the range of the 1 to several days. Similar result was also obtained by Burlaga and Viñas (2004) for lag of 1.3 days. Now, the PDF for $r_4 = 83.3$ days, presents a Gaussian format, where according Burlaga and Viñas (2004), this Gaussian behavior maybe showing a variety of flows characteristic of a particular epoch of the solar cycle activity.

Figure 4 presents also the theoretical and experimental semi logarithmic plots of $p_q(x_r)$ versus $x_r$ at same scales for proton density for 1997. We observe that the theoretical results (solid lines) are in good agreement with measurements across spatial scales spanning three orders of magnitude and a range of up to 5 standard deviations. Also, we performed the correlation coefficient between experimental and theoretical PDF for each scale, as shown in Table 2. Again, we can observer the higher values for correlation coefficient for all scales, showing good agreement between experimental data and our model. We can observe that all the PDFs exhibits stronger non-Gaussian behavior than the solar wind velocity PDFs. This distinct behavior between both the variables is due the peculiar characteristic of passive-scalar as pointed by Warhaft (2000) and Basu et al. (2003). This is an interesting aspect of differences between both the variables.

To study more closely this distinct behavior in both the variables, we also estimated the variation with scale of parameter q and plotted the parameter q by increment r for both quantities and for both the years, as shown in Figure 5. As a first analysis, we note that the four curves have similar behavior, where the parameter q value decreases as r grows. Katul et al. (1994) using a parameter related to scale kurtosis, the wavelet flatness factor (FF), have shown that in the inertial subrange scales, the more the separation distance r increases, the lower are the values of FF. They have also shown that this FF value trend is caused by the increase of intermittency in the smallest scales of the inertial subrange. In this sense, their results are similar to the ones presented in Figures 3 and 4 in which there is a clear enhancement of q as r decreases. The important aspect found is that the proton density data set is more intermittent than the solar wind velocity. This peculiar feature was observed in data set of passive scalars, like turbulent temperature in atmospheric of Amazonia (Bolzan et al., 2002; Ramos et al., 2004) and laboratory flows (Warhaft, 2000). Another interesting fact is that the intermittency level for both the variables is higher for 2000 if compared with 1997. This characteristic is important because it shows the difference between the two time-series for two different conditions: one time-series that represents a moderate solar activity period and, the other time-series that represents high solar activity.

From the analyses presented above, we note that the proton density time-series are more intermittent than the solar wind velocity. However, it is observed in Figure 3, we do not have good agreement between experimental and theoretical PDFs of the solar wind velocity. This behavior may be related of the particular skewness in solar wind velocity. Many investigations have reported the presence of the skweness in solar wind velocity time-series (Burlaga and Foreman, 2002; Burlaga and Viñas, 2004). Furthermore, Basu et al. (2003), through the proposed scheme to generate synthetic turbulent velocity and passive-scalar in hydrodynamics fields, show that in the small-scales the skewness presents the values of approximately –0.3 to –0.4 and, these small negative values are believed to be the origin of vortex stretching and nonlinear energy transfer from large to small scales. To show the skewness aspect, we performed the skewness for same four increment r to the solar wind velocity and proton density data, for both the years. Figure 6 displays only the skewness of the solar wind velocity, because the results for the proton density show the values of the skewness close to zero. The bulk velocity is usually between 200 and 700 km/s with an average of 400 km/s (Hargreaves, 1979). We note the high values in the skewness for 2000, in all increment scale r, if compared with 1997. We did not get negative values for the skewness, as obtained by Basu et al. (2003). This behavior may be indicating a different role of the skewness between the two turbulence, hydrodynamic and magnetohydrodynamic. However, we note that the solar activity has an important role in this parameter.

For understanding the influence of the solar activity in both the variables, we used an algorithm for Global Wavelet Spectrum (GWS) presented by Torrence and Compo (1998). This algorithm was applied in the solar wind and proton density time-series for two years, 1997 and 2000. Figure 7 shows the GWS for proton density for both the years. We note that there is increase of energy in some periods for 2000. In particular, the increase of energy with a period of approximately 26 days, corresponds to the solar rotation. We also observe the increase of the energy with lower periods like that 9 and 13 days. These periods can be explained through the fact that this year occurred many solar disturbances of short periods. For the solar wind velocity, the GWS show great differences between the both the years (Figure 8). Again, we note the increase of energy in all the periods for 2000, mainly in the short periods, corresponding to approximately 9 and 13 days. However, we did not observe influence of the solar activity in the increase of the energy in smaller periods of less than 1 day. This subject will be investigated in near future to understand how the energy transfer occurs between large and small scales during the high solar activity period of the Sun.

## Concluding Remarks

We have studied the PDFs of solar wind velocity and proton density for the two different periods of solar activity, 1997 (moderate activity) and 2000 (high activity), measured by spacecraft sensors. Our approach was based on generalized thermostatistics theory. The behavior of the entropic parameter q can be used to objectively quantify intermittency buildup in turbulent flows. From a practical point of view, the use of the entropic parameter as a measure of intermittency is justified by the fact that q is the key parameter that controls the shape of the PDF, which accurately models the statistics of turbulent solar wind and proton density. As expected from the earlier theoretical results, we found higher values of q in proton density for both the years. This is due to the peculiar characteristics of the scalar parameters. Similar results were obtained by Bolzan et al. (2002) and Ramos et al. (2004) using temperature time-series measured in the turbulent flow of the Amazonian forest. Among the physical mechanisms which would be responsible for this behaviour, we could mention the influence of the coherent magnetic vortices, studied by Kinney et al. (1995). As a consequence of this influence we did not observe good agreement between theoretical and experimental PDFs for the year 2000 solar wind velocity time series. Through the analysis of the skewness parameter related to solar wind time series, we observed high values of this parameter. The energy necessary to provide the increase in the skewness values was due to the increase of disturbances during this year. By Global Wavelet Spectrum (GWS) analyses, these disturbances increase the energy in lower periods for both the time-series in 2000, but being much efficient in the solar wind velocity time-series. These periods were approximately 9, 13 and 26 days. Taking into account these results we have shown that the generalized thermostatistics approach combining GWS analysis provides a simple and accurate framework for modeling the statistical behavior of MHD turbulence involved in the solar-terrestrial plasma dynamics.

## Acknowledgements

All the data used in this work were obtained from MTOF/PM Data by Carrington Rotation from the website: http://umtof.umd.edu/pm/crn/. Thanks are also due to the referees for their valuable suggestions and comments.

Figure Captions

Figure 1. Time-series for the year 1997. The top plot corresponds to the solar wind velocity and, the lower plot corresponds the proton density.

Figure 2: Same as in Figure 1 but for the year 2000.

Figure 3: Theoretical and experimental PDFs for the solar wind velocity in 1997.

Figure 4: Theoretical and experimental PDFs for the Proton Density in 1997.

Figure 5: Parameter q for different scales for both the variables and for the two years (1997 and 2000).

Figure 6: Increment scale variations of the skewness parameter for the solar wind velocity time-series.

Figure 7: Global Wavelet Spectrum (GWS) applied to the proton density time-series for the years 1997 and 2000.

Figure 8: Global Wavelet Spectrum (GWS) applied to the solar wind velocity time-series for the years 1997 and 2000.

| Increment r | Correlation Coefficient (%) |
|---|---|
| 2 | 92.81 |
| 20 | 94.03 |
| 200 | 98.15 |
| 2000 | 89.63 |

Table 1. Correlation coefficients between experimental and theoretical PDF to four increment scales for solar wind velocity time-series, 1997.

| Increment r | Correlation Coefficient (%) |
|---|---|
| 2 | 98.10 |
| 20 | 99.20 |
| 200 | 97.64 |
| 2000 | 99.04 |

Table 2. Correlation coefficients between experimental and theoretical PDF to four increment scales for proton density time-series, 1997.

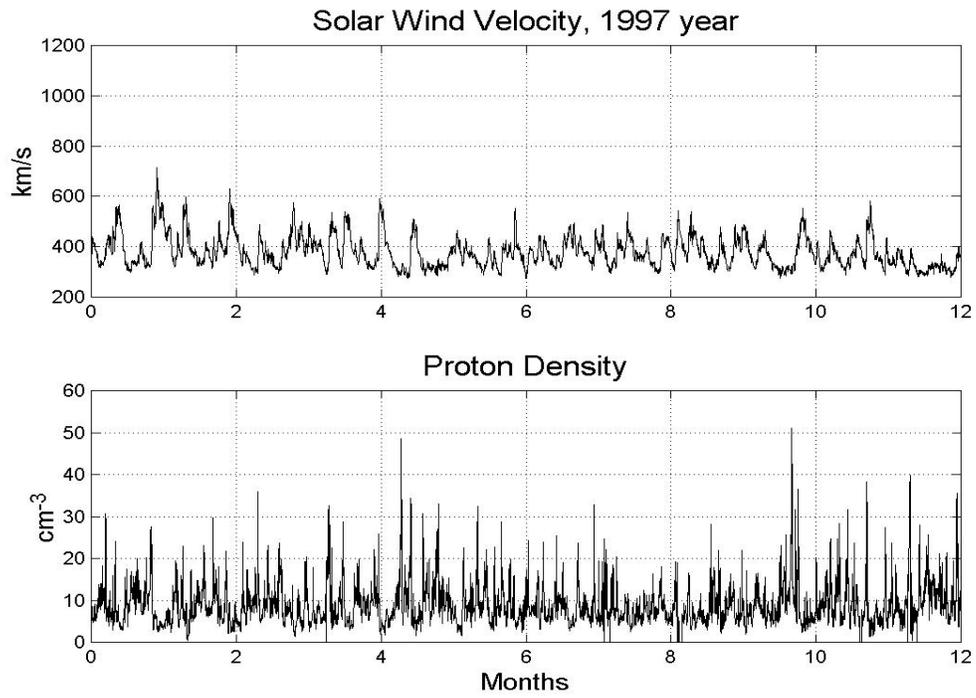

Figure 1: Time-series for the year 1997. The top plot corresponds to the solar wind velocity and, the lower plot corresponds the proton density.

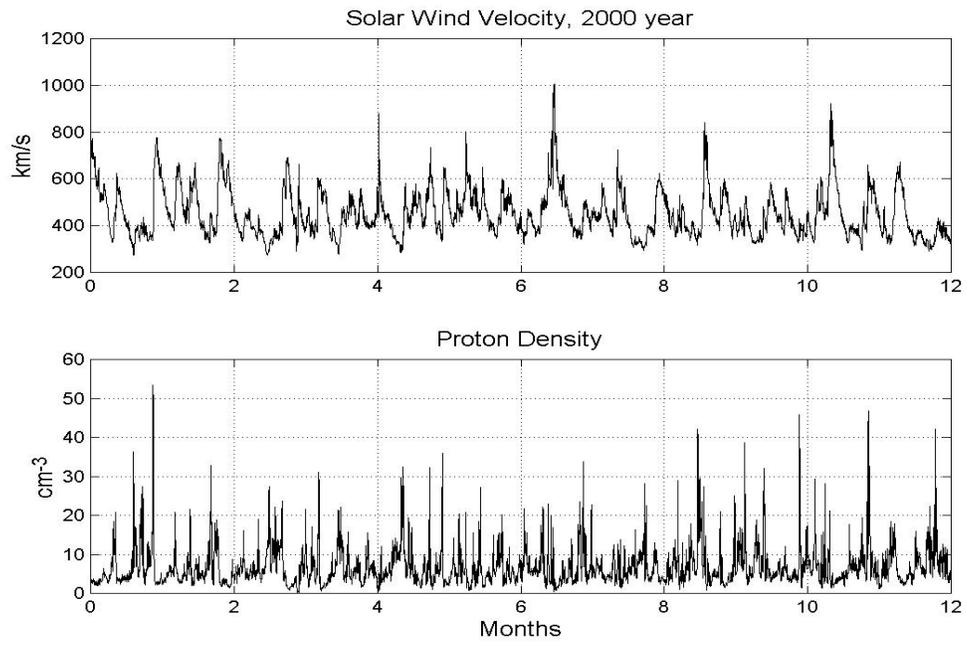

Figure 2: Same as in Figure 1 but for the year 2000.

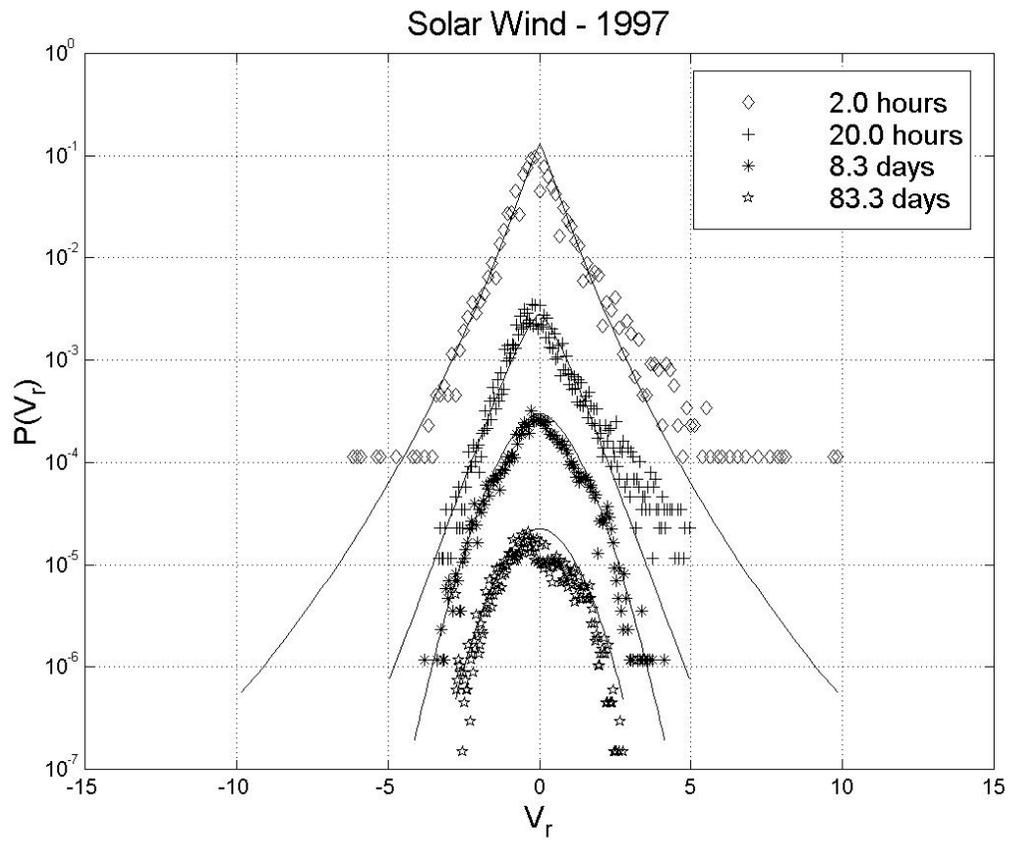

Figure 3: Theoretical and experimental PDFs for the solar wind velocity in 1997.

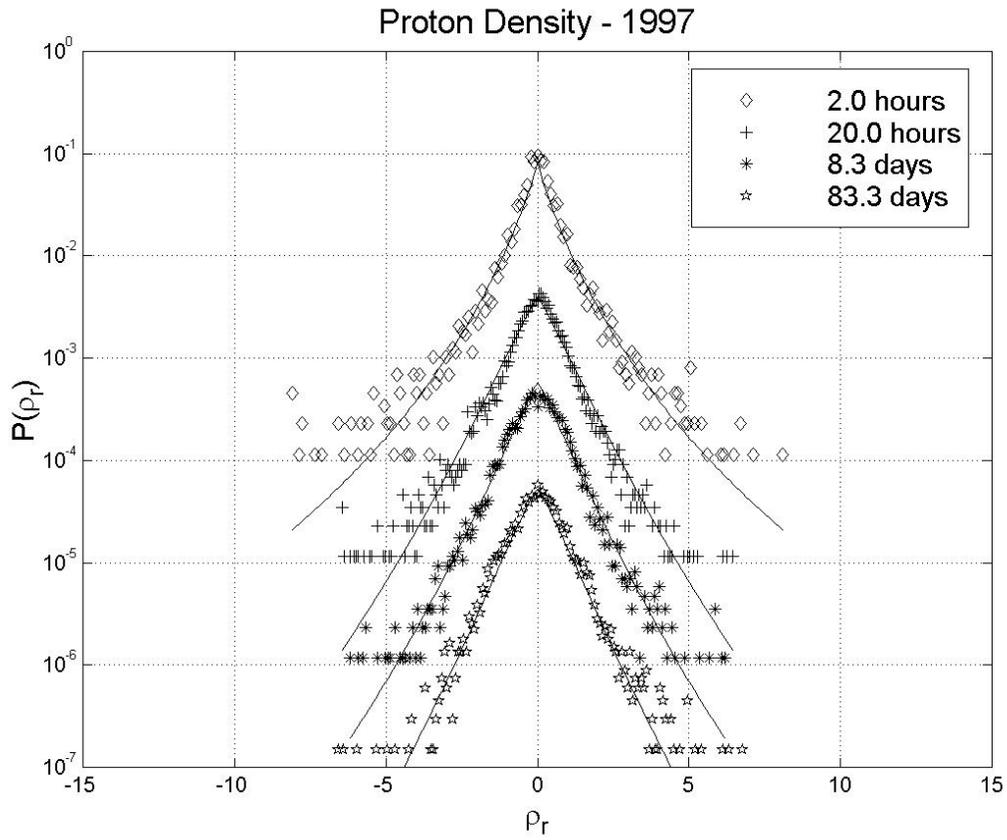

Figure 4: Theoretical and experimental PDFs for the Proton Density in 1997.

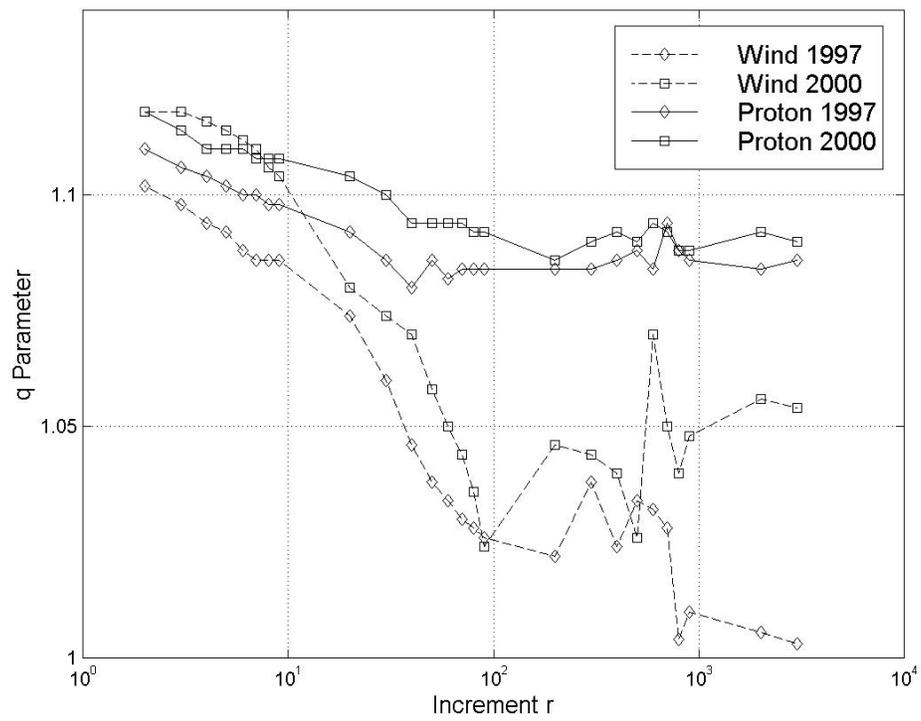

Figure 5: Parameter q for different scales for both the variables and for the two years (1997 and 2000).

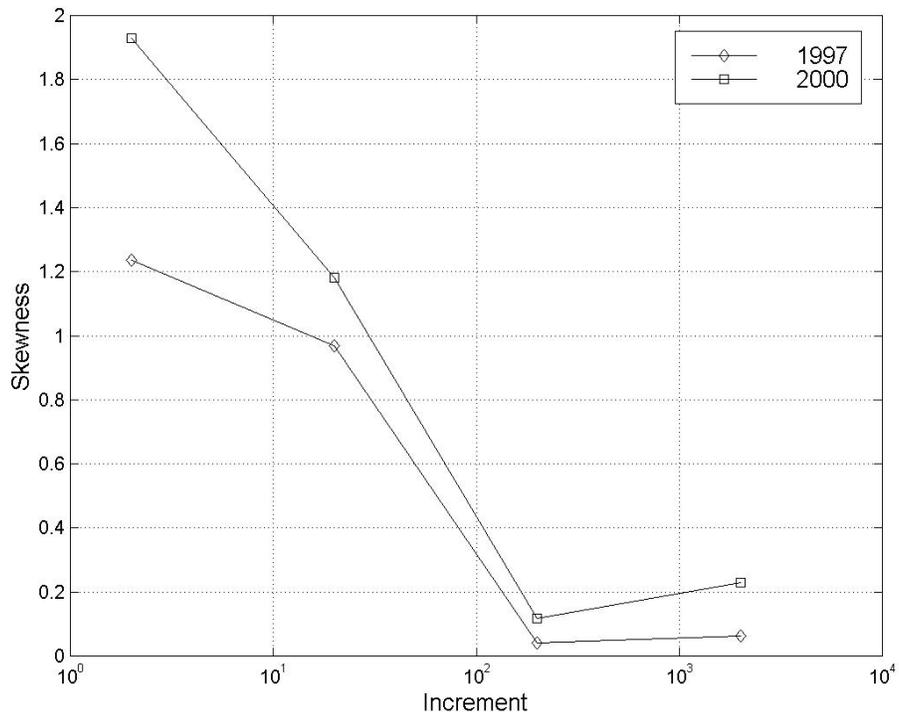

Figure 6: Increment scale variations of the skewness parameter for the solar wind velocity time-series.

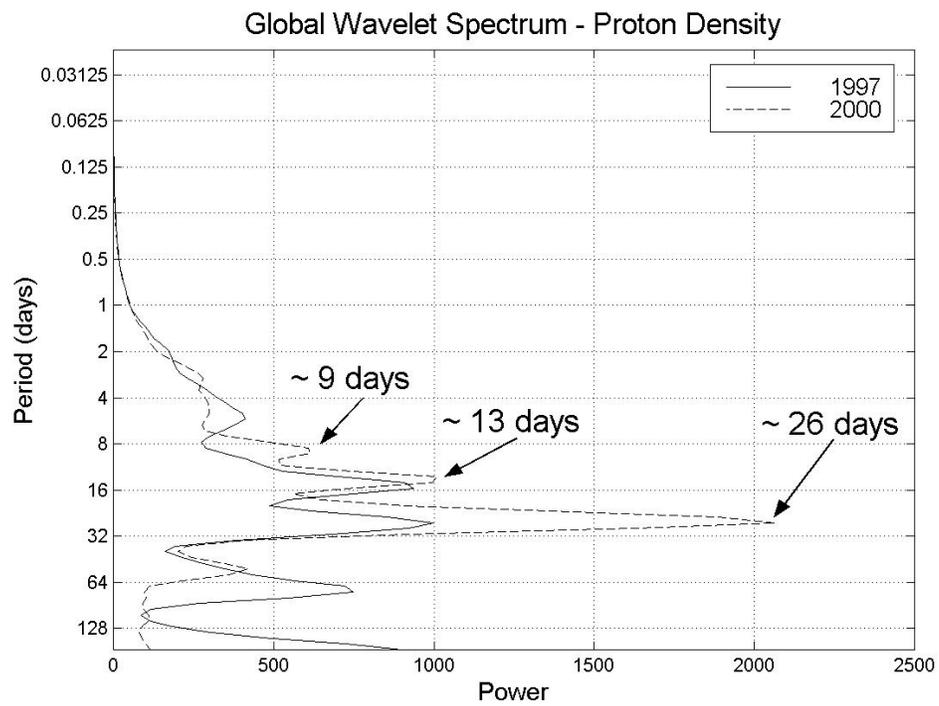

Figure 7: Global Wavelet Spectrum (GWS) applied to the proton density time-series for the years 1997 and 2000.

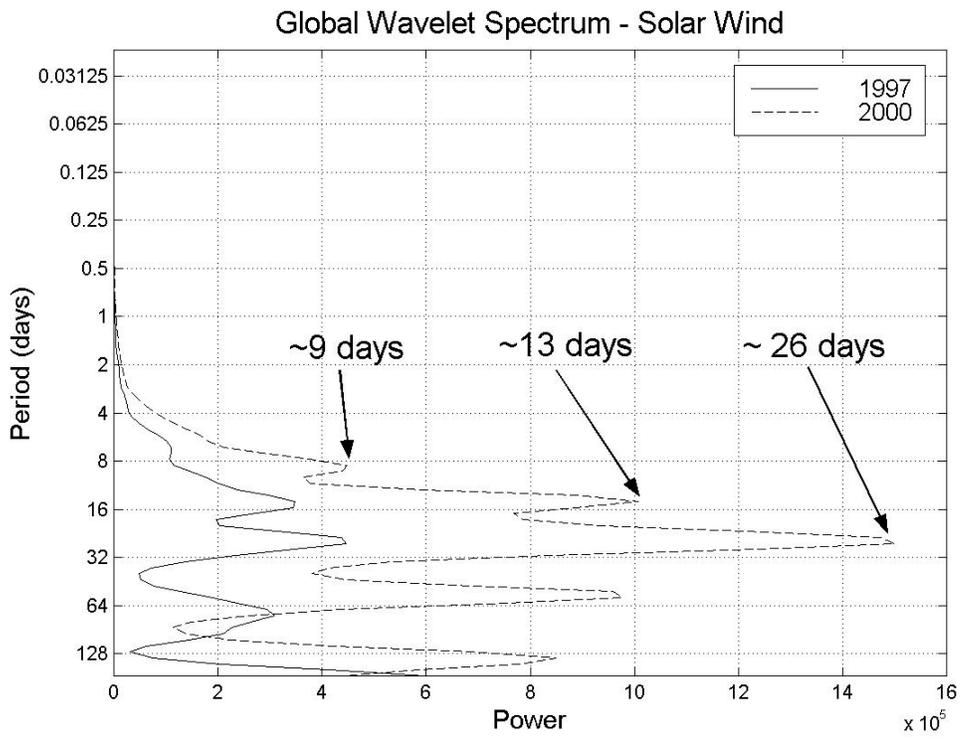

Figure 8: Global Wavelet Spectrum (GWS) applied to the solar wind velocity time-series for the years 1997 and 2000.